# Medium-entropy Engineering of magnetism in layered antiferromagnet $Cu_xNi_{2(1-x)}Cr_xP_2S_6$

Dinesh Upreti[1], Rabindra Basnet[1,2*], M. M. Sharma[1], Santosh Karki Chhetri[1], Gokul Acharya[1], Md Rafique Un Nabi[1,3], Josh Sakon[4], Mansour Mortazavi[2], Jin Hu[1,3,5*]

[1]Department of Physics, University of Arkansas, Fayetteville, Arkansas 72701, USA

[2]Department of Chemistry & Physics, University of Arkansas at Pine Bluff, Pine Bluff, Arkansas 71603, USA

[3]MonArk NSF Quantum Foundry, University of Arkansas, Fayetteville, Arkansas 72701, USA

[4]Department of Chemistry & Biochemistry, University of Arkansas, Fayetteville, Arkansas 72701, USA

[5]Materials Science and Engineering Program, Institute for Nanoscience and Engineering, University of Arkansas, Fayetteville, Arkansas, 72701, USA

Abstract

Engineering magnetism in layered magnets could result in novel phenomena related to two-dimensional (2D) magnetism, which can be useful for fundamental research and practical applications. Extensive doping efforts such as substitution and intercalation have been adopted to tune antiferromagnetic (AFM) properties in $M_2P_2X_6$ compounds. The substitutional doping in this material family has mainly focused on bimetallic substitution. Recently, the metal substitution can also be extended to more than two metal elements, leading to medium and high-entropy alloys (MEAs and HEAs), which are fairly underexplored in layered magnetic systems including $M_2P_2X_6$. In this work, we explored the magnetic properties of the previously unreported Cu- and Cr-substituted $Ni_2P_2S_6$ i.e., $Cu_xNi_{2(1-x)}Cr_xP_2S_6$. Our study reveals a relatively systematic evolution of AFM phases with substitution than that observed in traditional bimetallic substitution in $M_2P_2X_6$. Furthermore, the Cu and Cr substitutions in $Ni_2P_2S_6$ are found to enhance the ferromagnetic (FM) correlation, which is also accompanied by a possible weak FM phase at low temperatures for the intermediate compositions $0.32 \leq x \leq 0.80$. Our work provides a strategy to establish ferromagnetism in AFM $M_2P_2X_6$ that can also be used for property tuning in other layered magnets.

*jinhu@uark.edu

## I. Introduction

Tuning magnetic properties in layered magnets creates a pathway for a deeper understanding of magnetism in low dimensions and enlightens possible new routes to realize two-dimensional (2D) magnetic systems in real materials for spintronic applications [1–6]. Several strategies such as doping [7–33], high pressure [34–37] and electrostatic gating [3,38] have been implemented to tune magnetism in layered magnets. In addition, medium and high-entropy alloys (MEAs and HEAs), which are defined as a solid solution containing three to four and more than five principal elements respectively, can be an effective way to engineer magnetic properties. Owing to added complexcity and parameter space, such MEAs/HEAs are expected to generate unusual magnetic orderings arising from competing magnetic interactions enabled by expanded degrees of freedom [39].

Single crystalline medium or high entropy layered magnets are barely explored. Recently, a few medium [40] and high-entropy [41] layered antiferromagnetic (AFM) metal thiophosphates $MPX_3$ ($M$ = metal; $X$ = chalcogen S or Se) compounds have been reported. These studies have revealed the modulation of magnetism under medium/high entropy environment in $MPX_3$ [40,41], but detailed investigation is still lacking. For various $M$ and $X$, $MPX_3$ compounds exhibit common structural characteristics with metal atoms $M$ arranged in a honeycomb lattice and sandwiched by P and $X$ atoms, as shown in Fig. 1a. In this structure, P forms P-P dimers perpendicular to the hexagonal metal plane. In each dimer, each P is bonded with three $X$ to form $(P_2X_6)^{4-}$ bipyramids fill the center of the $M$ honeycomb lattice. Therefore, $MPX_3$ is also referred as $M_2P_2X_6$, which will be used throughout this article. The $M_2P_2X_6$ materials exhibit robust AFM ordering from bulk to atomically thin limit [1,6,16,18,20,42–60]. Tunable magnetism in this material family is mostly based on bi-metallic [i.e., $(M_I, M_{II})P_2X_6$, where $M_I$ and $M_{II}$ are different metals] [7–22,61] or bi-chalcogenide [i.e., $M_2P_2(S,Se)_6$] [26,27,60,62,63] substitutions, and inter-layer intercalations [30–33]. The $M$ and $X$ substitutions have resulted in diverse AFM structures [7–22,26,27,60,62,63] in addition to the signatures of ferrimagnetism [30,32,64] when doped with extra charge carriers by intercalating guest species. The isovalent substitution [65] and charge doping [66] have predicted ferromagnetism in $M_2P_2X_6$. So far, only Zn substitution in $Fe_2P_2S_6$ [67] and Co substitution in $Ni_2P_2S_6$ [21] have revealed the signatures of FM ordering. Even the reported medium/high entropy $M_2P_2X_6$ compounds lack ferromagnetism [40,41].

A new metal substitution strategy has been adopted in $M_2P_2X_6$, in which equal ratios of monovalent $M'^{1+}$ ($M$ = Cu or Ag) and trivalent $M''^{3+}$ ($M''$ = V or Cr) ions are fully substituted for $M^{2+}$ ion i.e., $(M^{2+})_2 = (M'^{+1}M''^{+3})$ to form a series of quaternary compounds $(Cu,Ag)(V,Cr)P_2X_6$ [68–76]. While these compounds have been discovered a long time ago [70,72–74], there have been limited magnetic studies [68–70,72–75] among which FM state has been observed in $CuCrP_2S_6$ [68,69] and $Ag(V,Cr)P_2Se_6$ [75,76]. In $CuCrP_2S_6$ [68,69] and $AgCrP_2Se_6$ [76], FM phase can be induced at around 6T field [68,69] that is much lower than the ultra-high field ($\mu_0 H \approx 35$ T) needed to achieve a fully polarized FM state in a monometallic $Fe_2P_2S_6$ [58]. Such AFM to FM transition at a relatively weaker field limit is commonly observed in Cr-based layered antiferromagnets [77–79], which is ascribed to a weak AFM coupling between $Cr^{3+}$ moments within an A-type magnetic ordering and a small magnetic anisotropy because of a lack of orbital degeneracy for $Cr^{3+}$ ($d^3$) ion [68,69,78,80,81]. Another compound $AgVP_2Se_6$ [75] exhibits a robust FM ground state from bulk to atomically thin limit, suggesting that the trivalent $Cr^{3+}$ and $V^{3+}$ ions may favor FM correlation in $M_2P_2X_6$.

Motivated by these works [68,69,75], here we report a new doping strategy in $M_2P_2X_6$, which involves the simultaneous mixing of mono, di, and trivalent metal ions. We successfully performed equal $Cu^{1+}$ and $Cr^{3+}$ substitutions for $Ni^{2+}$ in $Cu_xNi_{2(1-x)}Cr_xP_2S_6$ (or CNC) ($0.16 \le x \le 0.94$) leading to a wide range of medium-entropy compositions. Our work reveals a smoother evolution of AFM phases with substitution in CNC as compared to traditional bimetallic substitution in $M_2P_2X_6$. The intermediate CNC compositions $0.32 \le x \le 0.80$ reveal a possible weak FM phase at low temperatures, which is likely attributed to a net magnetic moment due to canted moments. The Cu and Cr (Cu+Cr) substitutions in $Ni_2P_2S_6$ enhance FM correlation, which is manifested by the field-driven moment polarization and systematic rise of saturation moment that is extremely challenging in $M_2P_2X_6$. Such rare medium-entropy engineering of magnetism provides a promising platform for property tuning in layered magnets.

## II. Experiment

The various $Cu_xNi_{2(1-x)}Cr_xP_2S_6$ ($0.16 \le x \le 0.94$) single crystals used in this work were synthesized by a chemical vapor transport method using $I_2$ as the transport agent. Elemental powders with desired ratios were sealed in a quartz tube and heated in a two-zone furnace with a temperature gradient from 750 to 550 °C for a week. The elemental compositions and crystal structures of the obtained crystals were examined by energy-dispersive x-ray spectroscopy (EDS) and powder x-ray diffraction (XRD) spectra obtained from grounded single crystals, respectively. Magnetization measurements were performed in physical and magnetic property measurement systems (PPMS and MPMS, Quantum Design).

## III. Results and discussion

Despite rapidly growing interest in Cr-based layered magnets because of their robust magnetic order down to 2D limit [82–86], the study of Cr-based $M_2P_2X_6$ compounds is surprisingly limited. So far, only $Cr_2P_2Se_6$ [87] and $Cr_{4/3}P_2S_6$ [88] have been reported while the stoichiometric $Cr_2P_2S_6$ has not been experimentally realized. The $Cr_2P_2S_6$ phase is difficult to stabilize due to a weaker Cr-S covalency, favoring 3+ valence for Cr instead of 2+ metal valence expected in $M_2P_2X_6$ [89]. With $Cr^{3+}$ seems more stable ionic state than $Cr^{2+}$, the direct substitution of Cr on metal sites in $M_2P_2X_6$ could be challenging and even produces a strong metal ion vacancy, as seen in $V_{2x}P_2S_6$ ($x$ = 0.78 [90,91] and 0.9 [92]) due to co-existence of $V^{2+}$ and $V^{3+}$ ions. This can be explained by only the highest 9% Cr substitution recently reported in $Ni_2P_2S_6$ [93], contrary to a much higher substitution of other metal elements like Mn, Fe, Co, Mg, etc. in $Ni_2P_2S_6$ [13,14,18–22,94]. The imbalance in charge valence created upon substituting $Cr^{3+}$ for $M^{2+}$ ions could be neutralized by replicating a strategy of co-substituting $Cr^{3+}$ and $M'^{1+}$ ($M'$ = Cu, Ag) ions in quaternary compounds (Cu,Ag)$CrP_2$(S,Se)$_6$ [68,69,71,76]. Following this idea, we were successful in substituting Cu and Cr in $Ni_2P_2S_6$. As shown in Fig. 1, our extensive CVT crystal growth has resulted in a wide range of single crystalline Cu, Cr, and Ni-substituted

CNC medium-entropy alloys. The nominal compositions used in the source materials and final compositions determined by EDS are presented in Table I. This reveals equal Cu and Cr substitutions for Ni in the grown crystals, which is expected and consistent with $CuCrP_2S_6$ [68,69] given the equivalent $Cu^{1+}$ and $Cr^{3+}$ ions are needed to maintain the charge neutrality. Therefore, these compositions can be represented as $Cu_xNi_{2(1-x)}Cr_xP_2S_6$, where $x$ represents Cu/Cr content.

The Cu and Cr substitutions in $Ni_2P_2S_6$ are further confirmed by the evolution of crystal structure through structure characterizations using x-ray diffraction (XRD). The XRD experiments were performed on powdered samples obtained by grinding single crystals. As shown in Fig. 1(b), although the XRD pattern for all $Cu_xNi_{2(1-x)}Cr_xP_2S_6$ samples look similar, our Rietveld refinement reveals two different monoclinic space groups $C2/m$ (Grey color) and $C2/c$ (Pink color) below and above $x = 0.46$ samples, which is understandable since monoclinic $C2/m$ and $C2/c$ structures have been determined for pristine $Ni_2P_2S_6$ and $CuCrP_2S_6$, respectively [68]. With increasing Cu and Cr substitutions in $x = 0$ (pristine $Ni_2P_2S_6$), a systematic low-angle peak shift is induced up to $x = 0.40$, which indicates an expansion of $Ni_2P_2S_6$ lattice and can be explained by the larger average size of $Cu^{1+}$ (0.96 Å) and $Cr^{3+}$ (0.62 Å) ions as compared to a $Ni^{2+}$ (0.70 Å) ion. The extracted lattice constants in Fig. 1(c) demonstrate a systematic increase in all three lattice parameters $a$, $b$, and $c$ upon Cu and Cr substitutions up to $x = 0.40$ in $Ni_2P_2S_6$. Further incorporating Cu and Cr causes a structural crossover to the monoclinic $CuCrP_2S_6$-type, which is accompanied by a clear XRD peak splitting [denoted by an asterisk in the right panel of Fig. 1(b)], consistent with the lowering symmetry from $C2/m$ to $C2/c$ space group. It is worth noting that the $x = 0.46$ sample displays a complicated XRD pattern due to the coexistence of both monoclinic phases as confirmed by our Rietveld refinement. The metal atoms in both $Ni_2P_2S_6$ and $CuCrP_2S_6$ can be seen in the honeycomb arrangement in the *ab*-plane when viewed along the *c*-axis [68,95] [Fig. 1(a)], which makes $C2/m$ and $C2/c$ crystal phases hardly distinguishable. However, in $CuCrP_2S_6$, the $Cr^{3+}$ ions are located almost in the center of each layer, whereas the $Cu^{1+}$ ions are off-centered along the *c*-axis [68,69,95], as depicted in Fig. 1(a). As a result, the *c*-axis [$c$ = 13.360(8) Å] is nearly doubled in $CuCrP_2S_6$ as compared to that of $Ni_2P_2S_6$ [$c$ = 6.616(3) Å] [68,69,95]. Indeed, the lattice parameters $a$ and $b$ lack substantial changes during the structure transition above $x = 0.40$ while the *c*-axis elongates significantly from $c$ = 6.620(6) Å for $x = 0.40$ to $c$ = 13.315(5) Å for $x = 0.46$, indicating structural evolution from $C2/m$ to $C2/c$ monoclinic phase with Cu and Cr substitutions for Ni.

In $M_2P_2X_6$, the isovalent metal-substituted compounds [7–22] exhibit a random distribution of two substituted metal ions in the honeycomb network. On the other hand, $M'^{1+}$ and $M''^{3+}$ ions in quaternary $(M'^{1+}M''^{3+})_2P_2X_6$ compounds have been proposed to form systematic patterns because of weaker repulsive coulomb interactions and substantial size differences between mono and trivalent ions [71]. For example, $Ag^{1+}$ and (V or Cr)$^{3+}$ ions in $Ag(V,Cr)P_2X_6$ ($X$ = S and Se) [70,71,75,76] form zig-zag chains while alternating arrangements of $Cu^{1+}$ and $Cr^{3+}$ ions are seen in $CuCrP_2S_6$ [68]. Similar systematic ordering of $Cu^{1+}$, $Ni^{2+}$, and $Cr^{3+}$ ions in a microscopic scale can be expected in CNC studied in this work, and given that the magnetic exchange interactions are highly dependent on the distribution of metal ions, this could significantly impact the magnetic properties.

To investigate the evolution of magnetic properties in medium-entropy CNC, we have measured the temperature dependence of susceptibility ($\chi$) under in-plane ($H//ab$) and out-of-plane ($H\perp ab$) magnetic fields of $\mu_0H$ = 0.1 T. The two end compounds $Ni_2P_2S_6$ ($x$ = 0) [18–20,22,26,42,47] and $CuCrP_2S_6$ ($x$ = 1) [68,69,73,95,96] have been found to order in C-type zig-zag [47] and A-type [73,95] AFM structures below $T_N \approx 155$ K and 32 K (denoted by black arrows in Fig. 2) respectively with magnetic easy axes aligned along or close to the *ab*-plane in both compounds, as depicted in Fig. 4. Owing to their distinct magnetic structures, the magnetic exchange interactions are different in these compounds. In $Ni_2P_2S_6$, the dominant third nearest-neighbor interaction $J_3$ is AFM although a much weaker FM interactions (nearest-neighbor interaction $J_1$ and inter-layer interaction $J_c$) also exist [97,98]. On the other hand, despite the AFM ground state in $CuCrP_2S_6$, the magnetic order is governed by intra-layer FM interactions while inter-layer AFM interactions are significantly weak [95]. Such distinct magnetic interactions in these compounds are in line with their different nature for PM to AFM transition in susceptibility measurements. $Ni_2P_2S_6$ lacks a sharp $T_N$ peak due to the presence of a broad hump in susceptibility just above $T_N$ arising from the short-range magnetic correlation within the 2D plane [18,19,26,47]. In this case, the PM to AFM transition $T_N$ can be estimated by the rise of anisotropy between in-plane ($\chi_{//}$) and out-of-plane ($\chi_{\perp}$) susceptibility, where $\chi_{//}$ must be smaller than $\chi_{\perp}$ below $T_N$ due to its in-plane AFM order [18–20,22,26,47]. In $CuCrP_2S_6$ [68,69], such a broad hump immediately above $T_N$ is absent that might be ascribed to the presence of Cu atoms which breaks the short-range magnetic correlation between Cr moments, therefore the AFM state is characterized by a sharp transition with a more pronounced $T_N$ peak observed in $\chi_{//}$ than $\chi_{\perp}$, consistent with its in-plane easy axis [68,69].

Such dissimilar AFM transition in susceptibility can be adopted to distinguish magnetic phases in substituted samples from $x = 0.16$ to 0.94 between $Ni_2P_2S_6$ and $CuCrP_2S_6$. For compositions $x = 0.16$-0.40 close to $Ni_2P_2S_6$, $\chi_{//}$ and $\chi_{\perp}$ resemble $Ni_2P_2S_6$-type susceptibility except for a sharp upturn at low temperature below $T$<50 K (Fig. 2). Before such dramatic susceptibility enhancement, a subtle transition similar to pristine $Ni_2P_2S_6$ is seen in $\chi_{//}$ leading to a weak but clear anisotropy between $\chi_{//}$ and $\chi_{\perp}$ (denoted by black arrows) with $\chi_{//} < \chi_{\perp}$, which can be defined as $T_N$ for these samples. In contrast, further increasing the substitution above $x = 0.40$ lacks such a weak transition before the sudden rise in susceptibility but is followed by a sharp $T_N$ peak in $\chi_{//}$ that is reminiscent of that of pristine $CuCrP_2S_6$ [68,69], implying $CuCrP_2S_6$-type AFM phase for these samples. Hence, the two magnetic regimes below and above $x = 0.40$ could be categorized as $Ni_2P_2S_6$-type (C-type zig-zag) and $CuCrP_2S_6$-type (A-type AFM) AFM phases that are denoted by $AFM_1$ and $AFM_2$ respectively (Fig. 4).

The changing AFM structure significantly modifies the magnetic exchange interactions, which is clarified by the evolution of magnetic ordering temperature. The variation of $T_N$ with (Cu+Cr) substitution for Ni suggests the AFM phase transition between $x = 0.40$ and 0.46 in CNC. After verifying with multiple samples, the $T_N$ displays a non-monotonic composition dependence, as summarized in the phase diagram in Fig. 4. For $x = 0$ to 0.40 in the $AFM_1$ region, the $T_N$ decreases systematically with (Cu+Cr) substitution. A slight increment of Cu and Cr contents to $x = 0.46$ produces a strong $T_N$ reduction followed by a monotonic enhancement with increasing $x$ on the $AFM_2$ regime. The suppression of $T_N$ with substitution in both end compounds could be ascribed to magnetic frustration similar to bimetallic substitutions in $M_2P_2X_6$ [12,16–18,20]. However, the sudden decline in $T_N$ from $x = 0.40$ to 0.46 is unusual and never seen in previous $M$-substituted $M_2P_2X_6$ and thus might not be clearly understood in terms of frustration alone. Instead, the sharp drop in $T_N$ between $x = 0.40$ and 0.46 corresponds well with the phase separation between $AFM_1$ and $AFM_2$ phases, hence it might be caused by modified magnetic interactions originating from AFM structure change. This can be explained by the overall attenuation of in-plane AFM interaction driven by enhanced FM correlation due to the emergence of A-type AFM order above $x \geq 0.46$ where magnetic interactions within each layer are FM in nature. Direct experimental probes such as neutron scattering, μSR, etc., and theoretical efforts are needed to clarify the magnetic structure evolution.

The variation of AFM phases between $x = 0.40$ and 0.46 i.e., for almost equal amounts of Ni and (Cu+Cr) is completely different than the previous metal-substituted $M_2P_2X_6$. For example, in $(Ni_{1-x}Mn_x)_2P_2S_6$, the magnetic structure changes from in-plane zig-zag ($Ni_2P_2S_6$) to out-of-plane Néel-type ($Mn_2P_2S_6$) at around $x = 0.75$ [99]. Similarly, the magnetic structure switches from out-of-plane zig-zag ($Fe_2P_2Se_6$) to in-plane Néel-type ($Mn_2P_2Se_6$) above $x = 0.90$ in $(Fe_{1-x}Mn_x)_2P_2Se_6$ [100]. This behavior might be due to the random substitution of two metal ions that creates frustration among magnetic interactions [99,100], as a result, the magnetic correlation in such a disordered magnetic plane may be governed by the single-ion anisotropy ($A$) of metal ions. Therefore, the evolution of magnetic structure is mainly dominated by the strongly anisotropic metal ions, which can be seen in both $(Ni_{1-x}Mn_x)_2P_2S_6$ [99] and $(Fe_{1-x}Mn_x)_2P_2Se_6$ [100] where AFM structure changes near $x = 0.75$ and 0.9 that means closer to relatively isotropic end compounds $Mn_2P_2S_6$ and $Mn_2P_2Se_6$, respectively. Thus, the change in magnetic structure for almost equal substitution of Ni and (Cu+Cr) in CNC studied in this work implies relatively less severe or entirely lack of random metal ion substitution, suggesting a systematic arrangement of $Cu^{1+}$, $Ni^{2+}$, and $Cr^{3+}$ ions, which may relieve magnetic frustration and promote a smoother evolution of magnetism. Such chemical ordering in CNC is also supported by the relative change in $T_N$ magnitude with substitution. As shown in Fig. 4, in the $AFM_1$ region, the $T_N$ for $Ni_2P_2S_6$ changes around 42.7 % upon 40% (Cu + Cr) substitution. Likewise, the $T_N$ varies by 39.4 % when 40 % Ni is substituted for (Cu+Cr) in $CuCrP_2S_6$ leading to an overall change of 43.8 % in the $AFM_2$ regime. Such identical doping dependence of $T_N$ in both end compounds is surprising and sharply contrasting to that of bimetallic substitutions. In $(Ni_{1-}$

$_x Mn_x)_2P_2S_6$ [18,20,99] and $(Fe_{1-x}Mn_x)_2P_2Se_6$ [100], the $T_N$ for $Ni_2P_2S_6$ and $Fe_2P_2Se_6$ decreases significantly more than that of $Mn_2P_2S_6$ and $Mn_2P_2Se_6$ respectively until reaching a minimum value upon 50% substitution (or $x = 0.5$). These results highlight the distinct metal-substitution dependence of magnetism in medium entropy CNC compounds, which might have originated from the systematic chemical ordering of substituted metal ions.

Such a well-defined $Cu^{1+}$, $Ni^{2+}$, and $Cr^{3+}$ order could even be more influential in tuning magnetic interactions in compositions consisting of the high amount of all three metal ions that help to realize an ideal medium-entropy environment. Indeed, this is evident in the low-temperature susceptibility of intermediate compositions $0.32 \leq x \leq 0.80$. As shown in Fig. 3(a), both $\chi_{//}$ and $\chi_{\perp}$ for these samples exhibit clear irreversibility between zero-field cooling (ZFC) and field-cooling (FC) measurements at low temperatures below $T_N$ (solid triangles) whereas other samples including the end compounds lack such susceptibility behavior. The ZFC and FC splitting is more prominent in $\chi_{//}$ than $\chi_{\perp}$ which is understandable for/nearly in-plane moment orientation over the entire composition range. Such ZFC and FC irreversibility has been observed under applied magnetic field at low temperatures in pristine $Mn_2P_2S_6$ [101] and $Ni_2P_2S_6$ [20], which have been attributed to weak FM phase arising from field-induced moment reorientation in AFM sublattices. The weak FM state derived from canted AFM order could also be possible here in CNC compounds because the ZFC and FC irreversibility in both magnetic phase regimes is more apparent and occurs at a greater temperature in samples $x = 0.40$ and 0.46 for which the substantial number of moments could be canted because of the strong interplay between competing magnetic exchange and anisotropy given the presence of a very high and comparable amount of substituted metal ions i.e., (40-46)% of $(Cu^{1+}+Cr^{3+})$ and (54-60)% of $Ni^{2+}$ ions. Hence, the irreversibility between ZFC and FC might be attributed to the FM component from the uncompensated canted moments. The onset of the weak FM phase for each composition is determined by the temperature when ZFC and FC splitting occurs [solid triangles in Fig. 3(a)], which is referred to as weak FM1 ($Weak_{FM1}$) and weak FM2 ($Weak_{FM2}$) corresponding to canted moments within C-type zig-zag and A-type AFM sublattices respectively, as depicted in the phase diagram in Fig. 4.

The irreversibility between ZFC and FC susceptibility is also seen for the spin-glass (SG) state. In $M_2P_2X_6$, the spin-glass transition has been observed upon metal substitution [12,40], which can be clarified by the frequency dependence of AC susceptibility. Because of the slow spin dynamics for an SG system, the spin relaxation time becomes longer. When an external AC magnetic field with a driving frequency is applied, the spin dynamics are enhanced with the increasing frequency of the AC field, consequently raising the SG temperature ($T_{SG}$). Here, the AC susceptibility for $x$ = 0.40 and 0.46 samples under varying frequencies from 100 to 10,000 Hz reveal the frequency-independent $T_{SG}$ [Fig. 3(b)]. These results are in stark contrast to the strong frequency dependence of AC susceptibility under a similar frequency range for SG state in isovalent bimetallic [12] and medium-entropy [40] $M_2P_2X_6$ compounds. This indicates that the SG state is absent in CNC and therefore the ZFC and FC irreversibility can be ascribed to a weak FM state due to canted moments. Indeed, the lack of an SG state seems plausible given the systematic arrangement of metal ions that could be immune from spin freezing.

The emergence of a weak FM state at low temperatures is further illustrated by the field dependence of magnetization at $T$ = 2 K [Fig. 5(a)]. In the $AFM_1$ region, pristine $Ni_2P_2S_6$ displays a robust AFM ground state that is manifested by a linear field-dependent magnetization followed by a metamagnetic transition above a high in-plane field of $\mu_0H \approx 6$ T with no magnetization saturation up to 9 T [20] [Fig. 5(a)]. Such lack of magnetic saturation at 9 T is understandable because a sister compound $Fe_2P_2S_6$ requires a much higher field ($\mu_0 H \approx 35$ T) for moment polarization [58]. For (Cu+Cr) substituted sample $x$ = 0.16 belonging to $AFM_1$ phase, the $T$ = 2 K in-plane magnetization slightly deviates from linearity [denoted by the black arrow in the inset in Fig. 5(a)] but still lacks a clear saturation behavior. However, at $T$ = 2 K, further increasing substitution to $x$ = 0.32 triggers the $weak_{FM1}$ phase, and simultaneously a moment polarization behavior featuring sublinear magnetization starts to occur at higher fields. The magnetization saturation ($M_{sat}$) becomes more apparent after subtracting the linear background and attains a value of ~(0.28±0.02)$\mu_B$ per f.u., as shown by dashed lines in Fig. 5(a). On increasing (Cu+Cr) content to $x$ = 0.40, the $weak_{FM1}$ phase becomes stronger because the ZFC and FC irreversibility and temperature for AFM to weak FM phase transition are enhanced [Fig. 3(a)], which is accompanied by a greater $M_{sat}$ ~(0.38±0.03)$\mu_B$ per f.u. than $x$ = 0.32 sample. These results indicate that canted moments within the $weak_{FM1}$ phase from $x$ = 0.32 to 0.40 facilitate moment polarization.

The scenario becomes different above $x$ = 0.40 in the $weak_{FM2}$ regime, where the canted AFM order may not play a substantial role in moment saturation. Substituting (Cu+Cr) to $x$ = 0.46 switches to $weak_{FM2}$ phase and the $M_{sat}$ is enhanced more than two times as compared to $x$ = 0.40. Such sudden rise in $M_{sat}$ might be attributed to the emergence of A-type AFM order as discussed above that exhibits a stronger intra-layer FM correlation which is likely to facilitate moment polarization. Increasing $x$ from 0.40 to 0.46 pushes the AFM to weak FM phase transition temperature to a slightly higher value, indicating the stronger weak FM phase which could also be the reason behind the sudden rise of $M_{sat}$. However, as seen in Fig. 5(b), the weak FM2 state is weakened with a further increment of (Cu+Cr) content until completely suppressed beyond $x$ = 0.80. The $M_{sat}$ on the other hand systematically increases with (Cu+Cr) substitution from ~$(0.86\pm0.04)\mu_B$ for $x$ = 0.46 to $(2.79\pm0.37)\mu_B$ for $x$ = 1. Such distinct composition dependence between the weak$_{FM2}$ phase and $M_{sat}$ suggests that the high-field moment saturation may not be strongly dependent on the FM component due to the canted moment but might be driven by the A-type AFM ground state. In fact, unlike $Ni_2P_2S_6$, the moment saturation seen above $\mu_0H \approx (6.56\pm0.20)$ T in $CuCrP_2S_6$ has been attributed to the weak inter-layer AFM interaction for A-type AFM ground state [68]. A similar situation can be expected in $x$ = 0.46 to 0.80 samples belonging to the weak$_{FM2}$ phase. Even if the moments are canted within an A-type AFM ground state, they might still be aligned ferromagnetically in each layer and bonded by relatively weaker exchange interactions between the layers like pristine $CuCrP_2S_6$. Thus, the moment saturation above $x \geq 0.46$ samples is likely ascribed to the formation of A-type AFM order. This also further suggests the AFM phase transition to an A-type AFM structure for $x \geq 0.46$.

As mentioned above, the $M_{sat}$ is found to attain a value of $(2.79\pm0.37)\mu_B$ at $T$ = 2 K in $CuCrP_2S_6$ that is close to the reported $M_{sat} \approx 3.00\mu_B$ per f.u. [68] and consistent with the theoretical saturation moment $(M_{sat})_{theo}$ expected for the $Cr^{3+}$ (S = 3/2) ion. The comparison of $M_{sat}$ value for all the samples with their corresponding $(M_{sat})_{theo} \approx (g_S S \mu_B)$, where $g_S$ ($\approx$ 2) is Landé $g$-factor and S = 1 and 3/2 for $Ni^{2+}$ and $Cr^{3+}$ respectively, are presented in Fig. 5(b). This demonstrates that the $M_{sat}$ deviates more from the theoretical value with increasing Ni substitution for (Cu+Cr) (or decreasing $x$ value). Substituting 6% Ni in $CuCrP_2S_6$ i.e, for $x$ = 0.94 reduces $M_{sat}$ by nearly 24% than that of theoretical value, which is surprising for such a low amount of Ni substitution and implies that each Ni moment prevents nearly three Cr moments from saturation. Further increasing the Ni content leads to the greater suppression of $M_{sat}$ in comparison to $(M_{sat})_{theo}$, reaching just 12% of the theoretical value (decreases by 88%). Such lower $M_{sat}$ than $(M_{sat})_{theo}$ indicates the lack of full spin polarization that is enhanced with increasing Ni substitution before lacking the moment saturation behavior up to 9 T field for $x < 0.32$ samples. The observed composition dependence of $M_{sat}$ can also be elucidated by the systematic increment of moment saturation field ($H_{sat}$) with (Cu+Cr) substitution [Inset; Fig. 5(b)]. With increasing (Cu+Cr) contents $x$, the number of saturated moments also increases, as a result, a higher magnetic field might be needed to polarize the moments, therefore the rise of $H_{sat}$ with (Cu+Cr) substitution is reasonable. Similar partial moment polarization has been recently reported in light 9% Cr-substituted $Ni_2P_2S_6$, which is attributed to the suppression of AFM interaction and magnetic anisotropy with Cr substitution [93]. A similar mechanism might occur in medium-entropy CNC compounds, demanding further in-depth experimental and theoretical investigations.

In addition to magnetism, the $Cu^{1+}$, $Ni^{2+}$, and $Cr^{3+}$ substituted CNC could be a promising material platform to study magneto-optics and magnetoelectric effects. One of the end compounds $Ni_2P_2S_6$ exhibits the strong coupling between magnetic order and light-matter interaction [59,102,103]. As discussed above, the (Cu+Cr) substitution in $Ni_2P_2S_6$ is highly efficient in tuning magnetic order and moment orientation, which could make unusual magneto-optical phenomena more accessible for novel photonic processes in layered magnets. Another end compound $CuCrP_2S_6$ displays interesting properties such as strong magnetoelectric coupling [95] and room-temperature ferroelectricity [104]. Thus, substituting Ni for (Cu+Cr) provides a large material pool for tunable magnetoelectric effect and ferroelectricity.

In conclusion, we studied the magnetic properties of Cu, Cr, and Ni substituted medium-entropy compounds $Cu_xNi_{2(1-x)}Cr_xP_2S_6$ and found a systematic tuning of magnetism with substitution. The (Cu+Cr) substitution for Ni induces a smooth evolution of AFM phases in contrast to magnetic frustration usually seen in conventional bimetallic substitutions. Increasing (Cu+Cr) contents also enhances FM correlation between magnetic moment, which is rare in previously reported $M_2P_2X_6$ compounds. These findings provide an interesting route to investigate tunable magnetism in layered magnets and extend the investigation to magneto-optics and magnetoelectric effects.

**Acknowledgments**

This work was primarily (crystal growth and intercalation) supported by the U.S. Department of Energy, Office of Science, Basic Energy Sciences program under Grant No. DE-SC0022006. We acknowledges the MonArk NSF Quantum Foundry for the magnetization measurements, which is supported by the National Science Foundation Q-AMASE-i program under NSF award No. DMR-1906383. R.B, M.M.S, and M.M acknowledges μ-ATOMS, an Energy Frontier Research Center funded by DOE, Office of Science, Basic Energy Sciences, under Award No. DE-SC0023412 (structure determination and part of the magnetic property analysis). J. S. acknowledges the support from NIH under award P20GM103429 for the powder XRD experiment.

## References

[1] G. Long, H. Henck, M. Gibertini, D. Dumcenco, Z. Wang, T. Taniguchi, K. Watanabe, E. Giannini, and A. F. Morpurgo, *Persistence of Magnetism in Atomically Thin MnPS3 Crystals*, Nano Lett. **20**, 2452 (2020).

[2] Z. Wang et al., *Electric-Field Control of Magnetism in a Few-Layered van Der Waals Ferromagnetic Semiconductor*, Nature Nanotechnology **13**, 7 (2018).

[3] B. Huang et al., *Electrical Control of 2D Magnetism in Bilayer CrI 3*, Nature Nanotechnology **13**, 7 (2018).

[4] Y. Wang et al., *Modulation Doping via a Two-Dimensional Atomic Crystalline Acceptor*, Nano Lett. **20**, 8446 (2020).

[5] A. R. C. McCray, Y. Li, R. Basnet, K. Pandey, J. Hu, D. P. Phelan, X. Ma, A. K. Petford-Long, and C. Phatak, *Thermal Hysteresis and Ordering Behavior of Magnetic Skyrmion Lattices*, Nano Lett. **22**, 7804 (2022).

[6] J.-U. Lee, S. Lee, J. H. Ryoo, S. Kang, T. Y. Kim, P. Kim, C.-H. Park, J.-G. Park, and H. Cheong, *Ising-Type Magnetic Ordering in Atomically Thin FePS3*, Nano Lett. **16**, 7433 (2016).

[7] N. Chandrasekharan and S. Vasudevan, *Dilution of a Layered Antiferromagnet: Magnetism in MnxZn1−xPS3*, Phys. Rev. B **54**, 14903 (1996).

[8] D. J. Goossens, A. J. Studer, S. J. Kennedy, and T. J. Hicks, *The Impact of Magnetic Dilution on Magnetic Order in MnPS3*, J. Phys.: Condens. Matter **12**, 4233 (2000).

[9] A. M. Mulders, J. C. P. Klaasse, D. J. Goossens, J. Chadwick, and T. J. Hicks, *High-Field Magnetization in the Diluted Quasi-Two-Dimensional Heisenberg Antiferromagnet Mn1-xZnxPS3*, J. Phys.: Condens. Matter **14**, 8697 (2002).

[10] Y. Takano, A. Arai, Y. Takahashi, K. Takase, and K. Sekizawa, *Magnetic Properties and Specific Heat of New Spin Glass Mn0.5Fe0.5PS3*, Journal of Applied Physics **93**, 8197 (2003).

[11] J. N. Graham, M. J. Coak, S. Son, E. Suard, J.-G. Park, L. Clark, and A. R. Wildes, *Local Nuclear and Magnetic Order in the Two-Dimensional Spin Glass Mn0.5Fe0.5PS3*, Phys. Rev. Materials **4**, 084401 (2020).

[12] T. Masubuchi, H. Hoya, T. Watanabe, Y. Takahashi, S. Ban, N. Ohkubo, K. Takase, and Y. Takano, *Phase Diagram, Magnetic Properties and Specific Heat of Mn1−xFexPS3*, Journal of Alloys and Compounds **460**, 668 (2008).

[13] V. Manríquez, P. Barahona, and O. Peña, *Physical Properties of the Cation-Mixed M′MPS3 Phases*, Materials Research Bulletin **35**, 1889 (2000).

[14] D. J. Goossens, S. Brazier-Hollins, D. R. James, W. D. Hutchison, and J. R. Hester, *Magnetic Structure and Glassiness in Fe0.5Ni0.5PS3*, Journal of Magnetism and Magnetic Materials **334**, 82 (2013).

[15] Y. He, Y.-D. Dai, H. Huang, J. Lin, and Y. Hsia, *The Ordering Distribution of the Metal Ions in the Layered Cation-Mixed Phosphorus Trisulfides MnxFe1−xPS3*, Journal of Alloys and Compounds **359**, 41 (2003).

[16] D. J. Goossens and T. J. Hicks, *The Magnetic Phase Diagram of MnxZn1-xPS3*, J. Phys.: Condens. Matter **10**, 7643 (1998).

[17] A. Bhutani, J. L. Zuo, R. D. McAuliffe, C. R. dela Cruz, and D. P. Shoemaker, *Strong Anisotropy in the Mixed Antiferromagnetic System Mn1−xFexPSe3*, Phys. Rev. Materials **4**, 034411 (2020).

[18] Y. Shemerliuk, Y. Zhou, Z. Yang, G. Cao, A. U. B. Wolter, B. Büchner, and S. Aswartham, *Tuning Magnetic and Transport Properties in Quasi-2D (Mn1−xNix)2P2S6 Single Crystals*, Electronic Materials **2**, 3 (2021).

[19] S. Selter, Y. Shemerliuk, M.-I. Sturza, A. U. B. Wolter, B. Büchner, and S. Aswartham, *Crystal Growth and Anisotropic Magnetic Properties of Quasi-Two-Dimensional (Fe1−xNix)2P2S6*, Phys. Rev. Materials **5**, 073401 (2021).

[20] R. Basnet, A. Wegner, K. Pandey, S. Storment, and J. Hu, *Highly Sensitive Spin-Flop Transition in Antiferromagnetic van Der Waals Material MPS3(M = Ni and Mn)*, Phys. Rev. Materials **5**, 064413 (2021).

[21] F. Wang et al., *Defect-Mediated Ferromagnetism in Correlated Two-Dimensional Transition Metal Phosphorus Trisulfides*, Science Advances **7**, eabj4086 (n.d.).

[22] S. Lee, J. Park, Y. Choi, K. Raju, W.-T. Chen, R. Sankar, and K.-Y. Choi, *Chemical Tuning of Magnetic Anisotropy and Correlations in Ni1−xFexPS3*, Phys. Rev. B **104**, 174412 (2021).

[23] C.-K. Tian, C. Wang, W. Ji, J.-C. Wang, T.-L. Xia, L. Wang, J.-J. Liu, H.-X. Zhang, and P. Cheng, *Domain Wall Pinning and Hard Magnetic Phase in Co-Doped Bulk Single Crystalline Fe3GeTe2*, Phys. Rev. B **99**, 184428 (2019).

[24] C. Tian, F. Pan, S. Xu, K. Ai, T. Xia, and P. Cheng, *Tunable Magnetic Properties in van Der Waals Crystals (Fe1−xCox)5GeTe2*, Appl. Phys. Lett. **116**, 202402 (2020).

[25] G. Drachuck, Z. Salman, M. W. Masters, V. Taufour, T. N. Lamichhane, Q. Lin, W. E. Straszheim, S. L. Bud'ko, and P. C. Canfield, *Effect of Nickel Substitution on Magnetism in the Layered van Der Waals Ferromagnet Fe3GeTe2*, Phys. Rev. B **98**, 144434 (2018).

[26] R. Basnet, K. M. Kotur, M. Rybak, C. Stephenson, S. Bishop, C. Autieri, M. Birowska, and J. Hu, *Controlling Magnetic Exchange and Anisotropy by Nonmagnetic Ligand Substitution in Layered MPX3 (M = Ni, Mn; X = S, Se)*, Phys. Rev. Research **4**, 023256 (2022).

[27] H. Han, H. Lin, W. Gan, Y. Liu, R. Xiao, L. Zhang, Y. Li, C. Zhang, and H. Li, *Emergent Mixed Antiferromagnetic State in MnPS3(1-x)Se3x*, Appl. Phys. Lett. **122**, 033101 (2023).

[28] M. Abramchuk, S. Jaszewski, K. R. Metz, G. B. Osterhoudt, Y. Wang, K. S. Burch, and F. Tafti, *Controlling Magnetic and Optical Properties of the van Der Waals Crystal CrCl3−xBrx via Mixed Halide Chemistry*, Advanced Materials **30**, 1801325 (2018).

[29] T. A. Tartaglia et al., *Accessing New Magnetic Regimes by Tuning the Ligand Spin-Orbit Coupling in van Der Waals Magnets*, Science Advances **6**, eabb9379 (n.d.).

[30] D. Tezze et al., *Tuning the Magnetic Properties of NiPS 3 through Organic-Ion Intercalation*, Nanoscale **14**, 1165 (2022).

[31] M. Mi et al., *Variation between Antiferromagnetism and Ferrimagnetism in NiPS3 by Electron Doping*, Advanced Functional Materials **32**, 2112750 (2022).

[32] R. Basnet, D. Ford, K. TenBarge, J. Lochala, and J. Hu, *Emergence of Ferrimagnetism in Li-Intercalated NiPS3*, J. Phys.: Condens. Matter **34**, 434002 (2022).

[33] D. Upreti, R. Basnet, M. M. Sharma, S. K. Chhetri, G. Acharya, M. R. U. Nabi, J. Sakon, B. Da, M. Mortazavi, and J. Hu, *Tuning Magnetism in Ising-Type van Der Waals Magnet FePS3 by Lithium Intercalation*, arXiv:2407.12662 (2024).

[34] T. Masubuchi, X. Jin, K. Koyama, Y. Takahashi, K. Takase, Y. Uwatoko, Y. Takano, and K. Sekizawa, *Magnetism and Interlayer Distance of MnPS3 Controlled by Intercalation and Applying High Pressure*, AIP Conference Proceedings **850**, 1279 (2006).

[35] T. Li et al., *Pressure-Controlled Interlayer Magnetism in Atomically Thin CrI3*, Nat. Mater. **18**, 12 (2019).

[36] S. Mondal, M. Kannan, M. Das, L. Govindaraj, R. Singha, B. Satpati, S. Arumugam, and P. Mandal, *Effect of Hydrostatic Pressure on Ferromagnetism in Two-Dimensional CrI3*, Phys. Rev. B **99**, 180407 (2019).

[37] Y. Peng et al., *Controlling Spin Orientation and Metamagnetic Transitions in Anisotropic van Der Waals Antiferromagnet CrPS4 by Hydrostatic Pressure*, Advanced Functional Materials **32**, 2106592 (2022).

[38] S. Jiang, L. Li, Z. Wang, K. F. Mak, and J. Shan, *Controlling Magnetism in 2D CrI 3 by Electrostatic Doping*, Nature Nanotechnology **13**, 7 (2018).

[39] L. Min, M. Sretenovic, T. W. Heitmann, T. W. Valentine, R. Zu, V. Gopalan, C. M. Rost, X. Ke, and Z. Mao, *A Topological Kagome Magnet in High Entropy Form*, Commun Phys **5**, 1 (2022).

[40] X. Chen, J. Wang, T. Ying, D. Huang, H. Gou, Q. Zhang, Y. Li, H. Hosono, J. Guo, and X. Chen, *Insulator-Metal-Superconductor Transition in the Medium-Entropy van Der Waals*

*Compound MPSe3 (M = Fe, Mn, Cd, and In) under High Pressure*, Phys. Rev. B **106**, 184502 (2022).

[41] T. Ying, T. Yu, Y.-S. Shiah, C. Li, J. Li, Y. Qi, and H. Hosono, *High-Entropy van Der Waals Materials Formed from Mixed Metal Dichalcogenides, Halides, and Phosphorus Trisulfides*, J. Am. Chem. Soc. **143**, 7042 (2021).

[42] K. Kim, S. Y. Lim, J.-U. Lee, S. Lee, T. Y. Kim, K. Park, G. S. Jeon, C.-H. Park, J.-G. Park, and H. Cheong, *Suppression of Magnetic Ordering in XXZ-Type Antiferromagnetic Monolayer NiPS 3*, Nature Communications **10**, 1 (2019).

[43] K. Kim et al., *Antiferromagnetic Ordering in van Der Waals 2D Magnetic Material MnPS3 Probed by Raman Spectroscopy*, 2D Mater. **6**, 041001 (2019).

[44] Z. Ni, A. V. Haglund, H. Wang, B. Xu, C. Bernhard, D. G. Mandrus, X. Qian, E. J. Mele, C. L. Kane, and L. Wu, *Imaging the Néel Vector Switching in the Monolayer Antiferromagnet MnPSe3 with Strain-Controlled Ising Order*, Nat. Nanotechnol. **16**, 7 (2021).

[45] F. Wang et al., *New Frontiers on van Der Waals Layered Metal Phosphorous Trichalcogenides*, Advanced Functional Materials **28**, 1802151 (2018).

[46] P. A. Joy and S. Vasudevan, *Magnetism in the Layered Transition-Metal Thiophosphates MPS3 (M=Mn, Fe, and Ni)*, Phys. Rev. B **46**, 5425 (1992).

[47] A. R. Wildes, V. Simonet, E. Ressouche, G. J. Mcintyre, M. Avdeev, E. Suard, S. A. Kimber, D. Lançon, G. Pepe, and B. Moubaraki, *Magnetic Structure of the Quasi-Two-Dimensional Antiferromagnet NiPS 3*, Physical Review B **92**, 224408 (2015).

[48] K. C. Rule, G. J. McIntyre, S. J. Kennedy, and T. J. Hicks, *Single-Crystal and Powder Neutron Diffraction Experiments on FePS3: Search for the Magnetic Structure*, Phys. Rev. B **76**, 134402 (2007).

[49] A. R. Wildes, H. M. Rønnow, B. Roessli, M. J. Harris, and K. W. Godfrey, *Static and Dynamic Critical Properties of the Quasi-Two-Dimensional Antiferromagnet MnPS3*, Phys. Rev. B **74**, 094422 (2006).

[50] D. Lançon, H. C. Walker, E. Ressouche, B. Ouladdiaf, K. C. Rule, G. J. McIntyre, T. J. Hicks, H. M. Rønnow, and A. R. Wildes, *Magnetic Structure and Magnon Dynamics of the Quasi-Two-Dimensional Antiferromagnet FePS3*, Phys. Rev. B **94**, 214407 (2016).

[51] A. R. Wildes, B. Roessli, B. Lebech, and K. W. Godfrey, *Spin Waves and the Critical Behaviour of the Magnetization in MnPS3*, J. Phys.: Condens. Matter **10**, 6417 (1998).

[52] A. R. Wildes, V. Simonet, E. Ressouche, R. Ballou, and G. J. McIntyre, *The Magnetic Properties and Structure of the Quasi-Two-Dimensional Antiferromagnet CoPS3*, J. Phys.: Condens. Matter **29**, 455801 (2017).

[53] T. Sekine, M. Jouanne, C. Julien, and M. Balkanski, *Light-Scattering Study of Dynamical Behavior of Antiferromagnetic Spins in the Layered Magnetic Semiconductor FePS3*, Phys. Rev. B **42**, 8382 (1990).

[54] Y. Takano, N. Arai, A. Arai, Y. Takahashi, K. Takase, and K. Sekizawa, *Magnetic Properties and Specific Heat of MPS3 (M=Mn, Fe, Zn)*, Journal of Magnetism and Magnetic Materials **272–276**, E593 (2004).

[55] G. Long et al., *Isolation and Characterization of Few-Layer Manganese Thiophosphite*, ACS Nano **11**, 11330 (2017).

[56] K. Okuda, K. Kurosawa, S. Saito, M. Honda, Z. Yu, and M. Date, *Magnetic Properties of Layered Compound MnPS3*, J. Phys. Soc. Jpn. **55**, 4456 (1986).

[57] D. J. Goossens, A. R. Wildes, C. Ritter, and T. J. Hicks, *Ordering and the Nature of the Spin Flop Phase Transition in MnPS3*, J. Phys.: Condens. Matter **12**, 1845 (2000).

[58] A. R. Wildes, D. Lançon, M. K. Chan, F. Weickert, N. Harrison, V. Simonet, M. E. Zhitomirsky, M. V. Gvozdikova, T. Ziman, and H. M. Rønnow, *High Field Magnetization of FePS3*, Phys. Rev. B **101**, 024415 (2020).

[59] D. Afanasiev et al., *Controlling the Anisotropy of a van Der Waals Antiferromagnet with Light*, Science Advances **7**, eabf3096 (n.d.).

[60] A. Wiedenmann, J. Rossat-Mignod, A. Louisy, R. Brec, and J. Rouxel, *Neutron Diffraction Study of the Layered Compounds MnPSe3 and FePSe3*, Solid State Communications **40**, 1067 (1981).

[61] J. P. Odile, J. J. Steger, and Aaron. Wold, *Preparation and Properties of the Solid Solution Series Zinc Iron Phosphorus Trisulfide (Zn1-xFexPS3)*, Inorg. Chem. **14**, 2400 (1975).

[62] S. Calder, A. V. Haglund, A. I. Kolesnikov, and D. Mandrus, *Magnetic Exchange Interactions in the van Der Waals Layered Antiferromagnet MnPSe3*, Phys. Rev. B **103**, 024414 (2021).

[63] P. Jeevanandam and S. Vasudevan, *Magnetism in MnPSe3: A Layered 3d5 Antiferromagnet with Unusually Large XY Anisotropy*, J. Phys.: Condens. Matter **11**, 3563 (1999).

[64] M. Mi et al., *Variation between Antiferromagnetism and Ferrimagnetism in NiPS3 by Electron Doping*, arXiv:2112.02544 [Cond-Mat] (2022).

[65] J. Yang, Y. Zhou, Y. Dedkov, and E. Voloshina, *Dirac Fermions in Half-Metallic Ferromagnetic Mixed Cr1−xMxPSe3 Monolayers*, Advanced Theory and Simulations **3**, 2000228 (2020).

[66] B. L. Chittari, Y. Park, D. Lee, M. Han, A. H. MacDonald, E. Hwang, and J. Jung, *Electronic and Magnetic Properties of Single-Layer MPX3 Metal Phosphorous Trichalcogenides*, Phys. Rev. B **94**, 184428 (2016).

[67] J. Peng et al., *Ferromagnetism Induced by Magnetic Dilution in Van Der Waals Material Metal Thiophosphates*, Advanced Quantum Technologies **6**, 2200105 (2023).

[68] S. Selter et al., *Crystal Growth, Exfoliation, and Magnetic Properties of Quaternary Quasi-Two-Dimensional CuCrP2S6*, Phys. Rev. Mater. **7**, 033402 (2023).

[69] X. Wang et al., *Electrical and Magnetic Anisotropies in van Der Waals Multiferroic CuCrP2S6*, Nat Commun **14**, 1 (2023).

[70] C. Payen, H. Mutka, J. L. Soubeyroux, P. Molinié, and P. Colombet, *Static and Dynamic Properties of the Quasi-1D Heisenberg Antiferromagnets AgVP2S6 (S=1) and AgCrP2S6 (S = 3/2)*, Journal of Magnetism and Magnetic Materials **104–107**, 797 (1992).

[71] S. Selter, Y. Shemerliuk, B. Büchner, and S. Aswartham, *Crystal Growth of the Quasi-2D Quarternary Compound AgCrP2S6 by Chemical Vapor Transport*, Crystals **11**, 5 (2021).

[72] H. Mutka, C. Payen, P. Molinié, J. L. Soubeyroux, P. Colombet, and A. D. Taylor, *Dynamic Structure Factor [S(Q,ω)] of the S=1 Quasi-One-Dimensional Heisenberg Antiferromagnet: Neutron-Scattering Study on AgVP2S6*, Phys. Rev. Lett. **67**, 497 (1991).

[73] P. Colombet, A. Leblanc, M. Danot, and J. Rouxel, *Structural Aspects and Magnetic Properties of the Lamellar Compound Cu0.50Cr0.50PS3*, Journal of Solid State Chemistry **41**, 174 (1982).

[74] S. Lee, P. Colombet, G. Ouvrard, and R. Brec, *A New Chain Compound of Vanadium (III): Ag12V12PS3 Structure, Metal Ordering, and Magnetic Properties*, Materials Research Bulletin **21**, 917 (1986).

[75] Y. Peng et al., *A Quaternary van Der Waals Ferromagnetic Semiconductor AgVP2Se6*, Advanced Functional Materials **30**, 1910036 (2020).

[76] M. A. Susner et al., *Structural, Magnetic, and Optical Properties of the van Der Waals Antiferromagnet AgCrP2Se6*, J. Phys. Chem. C (2024).

[77] M. A. McGuire, G. Clark, S. KC, W. M. Chance, G. E. Jellison, V. R. Cooper, X. Xu, and B. C. Sales, *Magnetic Behavior and Spin-Lattice Coupling in Cleavable van Der Waals Layered CrCl3 Crystals*, Phys. Rev. Mater. **1**, 014001 (2017).

[78] Y. Peng et al., *Magnetic Structure and Metamagnetic Transitions in the van Der Waals Antiferromagnet CrPS4*, Advanced Materials **32**, 2001200 (2020).

[79] E. J. Telford et al., *Layered Antiferromagnetism Induces Large Negative Magnetoresistance in the van Der Waals Semiconductor CrSBr*, Advanced Materials **32**, 2003240 (2020).

[80] B. Kuhlow, *Magnetic Ordering in CrCl3 at the Phase Transition*, Physica Status Solidi (a) **72**, 161 (1982).

[81] J. W. Cable, M. K. Wilkinson, and E. O. Wollan, *Neutron Diffraction Investigation of Antiferromagnetism in CrCl3*, Journal of Physics and Chemistry of Solids **19**, 29 (1961).

[82] B. Huang et al., *Layer-Dependent Ferromagnetism in a van Der Waals Crystal down to the Monolayer Limit*, Nature **546**, 7657 (2017).

[83] C. Gong et al., *Discovery of Intrinsic Ferromagnetism in Two-Dimensional van Der Waals Crystals*, Nature **546**, 265 (2017).

[84] A. Bedoya-Pinto, J.-R. Ji, A. K. Pandeya, P. Gargiani, M. Valvidares, P. Sessi, J. M. Taylor, F. Radu, K. Chang, and S. S. P. Parkin, *Intrinsic 2D-XY Ferromagnetism in a van Der Waals Monolayer*, Science **374**, 616 (2021).

[85] X. Cai et al., *Atomically Thin CrCl3: An In-Plane Layered Antiferromagnetic Insulator*, Nano Lett. **19**, 3993 (2019).

[86] J. Son et al., *Air-Stable and Layer-Dependent Ferromagnetism in Atomically Thin van Der Waals CrPS4*, ACS Nano **15**, 16904 (2021).

[87] M. Baithi et al., *Incommensurate Antiferromagnetic Order in Weakly Frustrated Two-Dimensional van Der Waals Insulator CrPSe3*, Inorg. Chem. **62**, 12674 (2023).

[88] Z.-L. Huang, J.-T. Zhao, J.-X. Mi, S.-Y. Mao, and L.-S. Zheng, *Room Temperature Solid State Synthesis and Characterization of a New Chromium Thiophosphate Cr4(P2S6)3*, Journal of Solid State Chemistry **144**, 388 (1999).

[89] D. Sen and T. Saha-Dasgupta, *Pressure-Tuned Valence Transition, Insulator-Metal Transition in van Der Waals Antiferromagnet CrPS3*, Phys. Rev. Mater. **7**, 064008 (2023).

[90] G. Ouvrard, R. Fréour, R. Brec, and J. Rouxel, *A Mixed Valence Compound in the Two Dimensional MPS3 Family: V0.78PS3 Structure and Physical Properties*, Materials Research Bulletin **20**, 1053 (1985).

[91] K. Ichimura and M. Sano, *Electrical Conductivity of Layered Transition-Metal Phosphorus Trisulfide Crystals*, Synthetic Metals **45**, 203 (1991).

[92] M. J. Coak, S. Son, D. Daisenberger, H. Hamidov, C. R. S. Haines, P. L. Alireza, A. R. Wildes, C. Liu, S. S. Saxena, and J.-G. Park, *Isostructural Mott Transition in 2D Honeycomb Antiferromagnet V0.9PS 3*, Npj Quantum Materials **4**, 1 (2019).

[93] R. Basnet, D. Upreti, T. Patel, S. K. Chhetri, G. Acharya, M. R. U. Nabi, M. M. Sharma, J. Sakon, M. Mortazavi, and J. Hu, *Field-Induced Spin Polarization in Lightly Cr-Substituted Layered Antiferromagnet NiPS3*, arXiv:2404.02091.

[94] D. J. Goossens, D. James, J. Dong, R. E. Whitfield, L. Norén, and R. L. Withers, *Local Order in Layered NiPS3 and Ni0.7Mg0.3PS3*, J. Phys.: Condens. Matter **23**, 065401 (2011).

[95] C. B. Park, A. Shahee, K.-T. Kim, D. R. Patil, S. A. Guda, N. Ter-Oganessian, and K. H. Kim, *Observation of Spin-Induced Ferroelectricity in a Layered van Der Waals Antiferromagnet CuCrP2S6*, Advanced Electronic Materials **8**, 2101072 (2022).

[96] V. Maisonneuve, C. Payen, and V. B. Cajipe, *On CuCrP2S6: Copper Disorder, Stacking Distortions, and Magnetic Ordering*, Journal of Solid State Chemistry **116**, 208 (1995).

[97] D. Lançon, R. A. Ewings, T. Guidi, F. Formisano, and A. R. Wildes, *Magnetic Exchange Parameters and Anisotropy of the Quasi-Two-Dimensional Antiferromagnet NiPS3*, Phys. Rev. B **98**, 134414 (2018).

[98] A. R. Wildes, J. R. Stewart, M. D. Le, R. A. Ewings, K. C. Rule, G. Deng, and K. Anand, *Magnetic Dynamics of NiPS3*, Phys. Rev. B **106**, 174422 (2022).

[99] Z. Lu, X. Yang, L. Huang, X. Chen, M. Liu, J. Peng, S. Dong, and J.-M. Liu, *Evolution of Magnetic Phase in Two-Dimensional van Der Waals Mn1−xNixPS3 Single Crystals*, J. Phys.: Condens. Matter **34**, 354005 (2022).

[100] R. Basnet, T. Patel, J. Wang, D. Upreti, S. K. Chhetri, G. Acharya, M. R. U. Nabi, J. Sakon, and J. Hu, *Understanding and Tuning Magnetism in Layered Ising-Type Antiferromagnet FePSe3 for Potential 2D Magnet*, Advanced Electronic Materials **n/a**, 2300738 (n.d.).

[101] H. Han et al., *Field-Induced Spin Reorientation in the Néel-Type Antiferromagnet MnPS3*, Phys. Rev. B **107**, 075423 (2023).

[102] S. Kang et al., *Coherent Many-Body Exciton in van Der Waals Antiferromagnet NiPS3*, Nature **583**, 7818 (2020).

[103] X. Wang et al., *Spin-Induced Linear Polarization of Photoluminescence in Antiferromagnetic van Der Waals Crystals*, Nat. Mater. **20**, 7 (2021).

[104] W. F. Io et al., *Direct Observation of Intrinsic Room-Temperature Ferroelectricity in 2D Layered CuCrP2S6*, Nat Commun **14**, 1 (2023).

# Figures

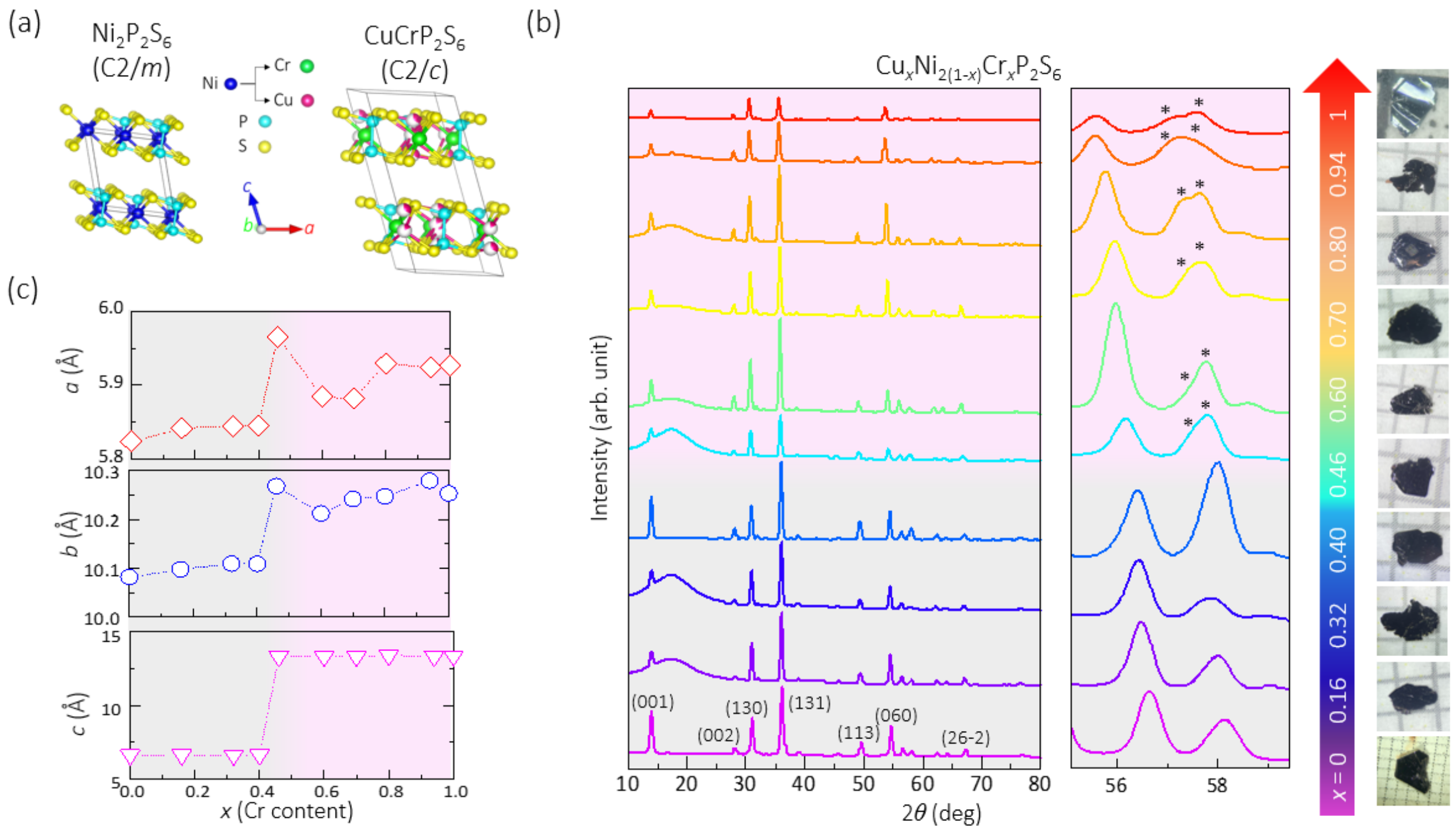

FIG. 1. (a) Crystal structures of $Ni_2P_2S_6$ and $CuCrP_2S_6$. (b) X-ray diffraction (XRD) results for Cu and Cr-substituted $Ni_2P_2S_6$, $Cu_xNi_{2(1-x)}Cr_xP_2S_6$ ($0 \leq x \leq 1$). Right panel shows the evolution of XRD peaks and the optical microscope images of single crystals of $Cu_xNi_{2(1-x)}Cr_xP_2S_6$. (c) Composition dependence of lattice parameters *a*, *b*, and *c* in $Cu_xNi_{2(1-x)}Cr_xP_2S_6$. The different colored regions in Figs. 1(b and c) represent different crystal structures for $Cu_xNi_{2(1-x)}Cr_xP_2S_6$ (grey: monoclinic space group C2/m; pink: monoclinic space group C2/*c*).

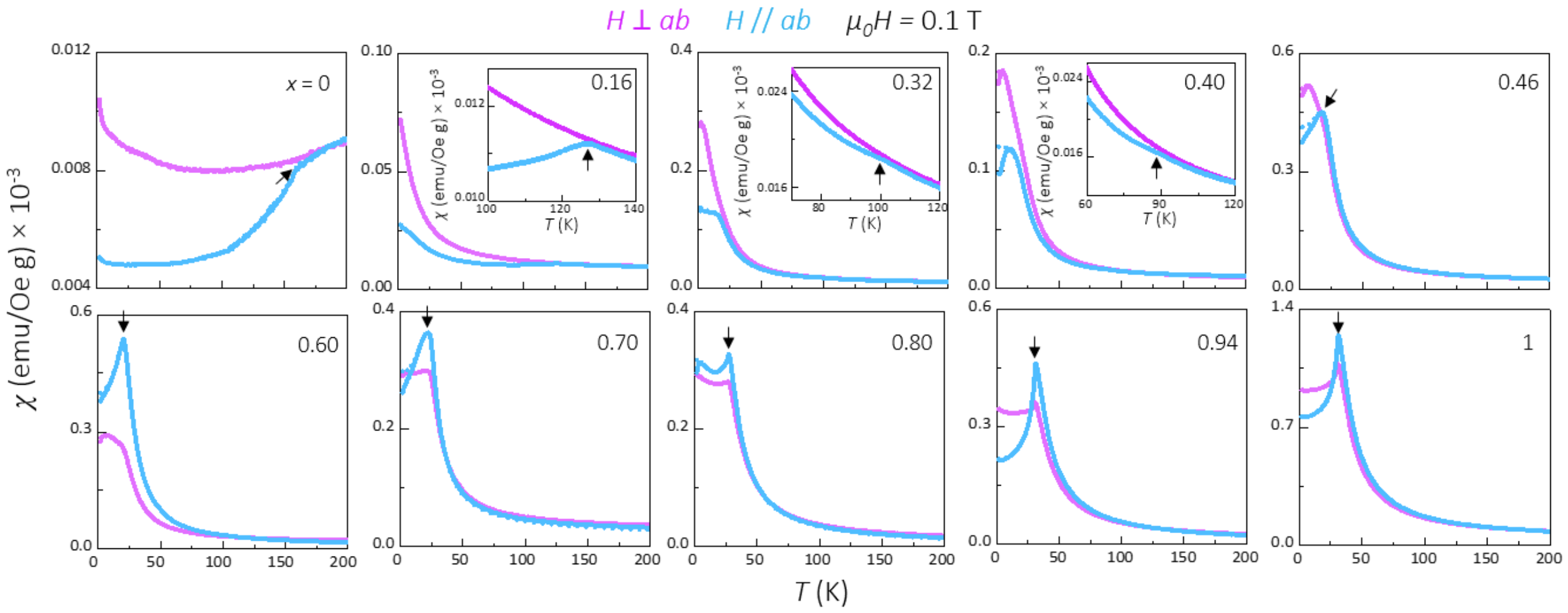

FIG. 2. Temperature dependence of susceptibility ($\chi$) for $Cu_xNi_{2(1-x)}Cr_xP_2S_6$ ($0 \leq x \leq 1$) samples under in-plane ($H||ab$, blue) and out-of-plane ($H\perp ab$, magenta) magnetic fields of $\mu_0 H$ = 0.1 T. Inset: Zoom-in of low temperature susceptibility to identify magnetic transitions. The black arrows denote the antiferromagnetic (AFM) transition temperature $T_N$.

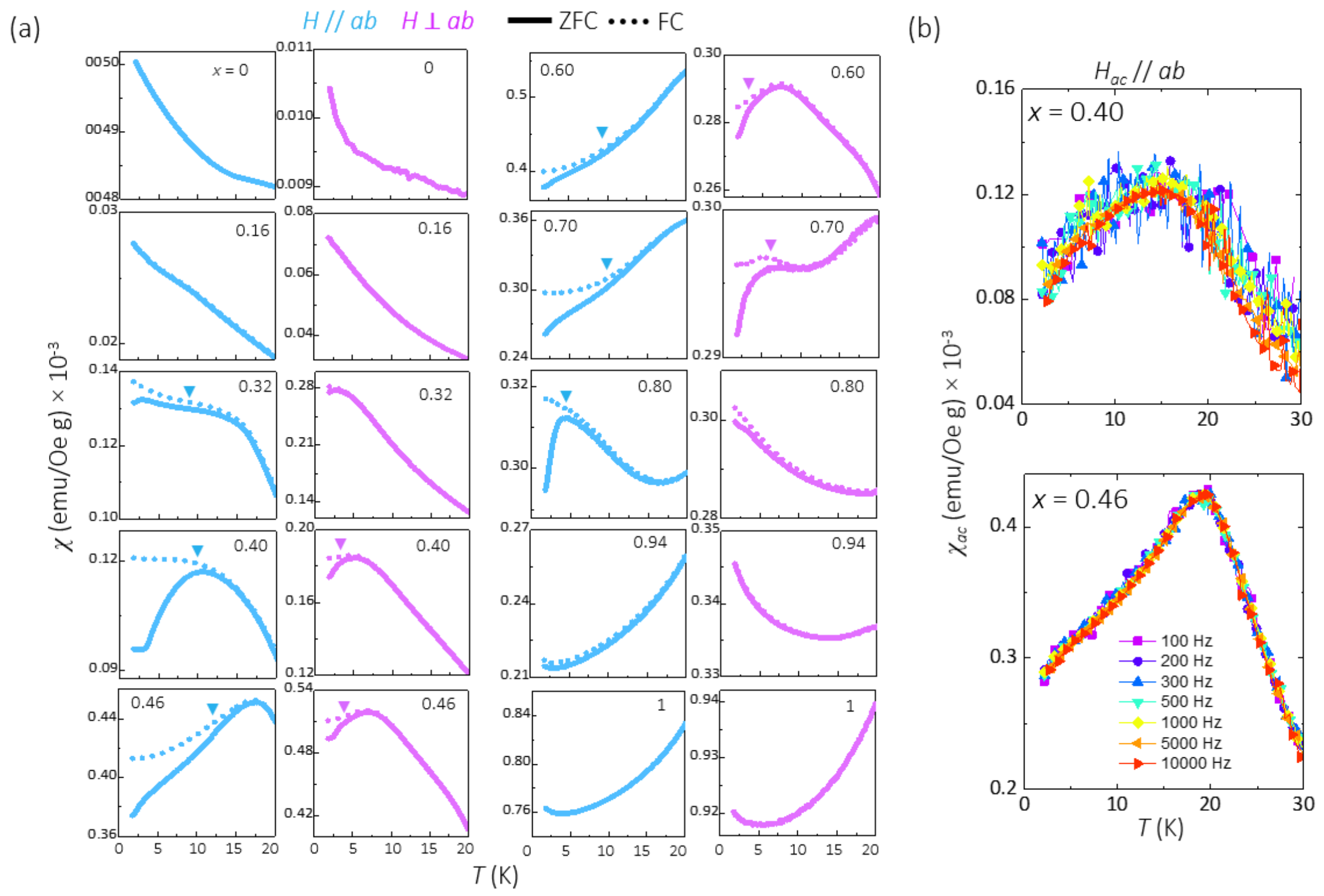

FIG. 3. (a) Low-temperature zero-field cooling (ZFC; solid lines) and field-cooling (FC; dashed lines) susceptibility ($\chi$) for $Cu_xNi_{2(1-x)}Cr_xP_2S_6$ ($0 \leq x \leq 1$) samples under in-plane ($H \| ab$, blue) and out-of-plane ($H \perp ab$, magenta) magnetic fields of $\mu_0 H$ = 0.1 T. The solid triangles denote the onset of ZFC and FC irreversibility. (b) Temperature dependence of AC-susceptibility ($\chi_{ac}$) for $x$ = 0.40 (upper panel) and 0.46 (lower panel) samples under an in-plane AC field of 10 Oe at different frequencies from 100 Hz to 10,000 Hz.

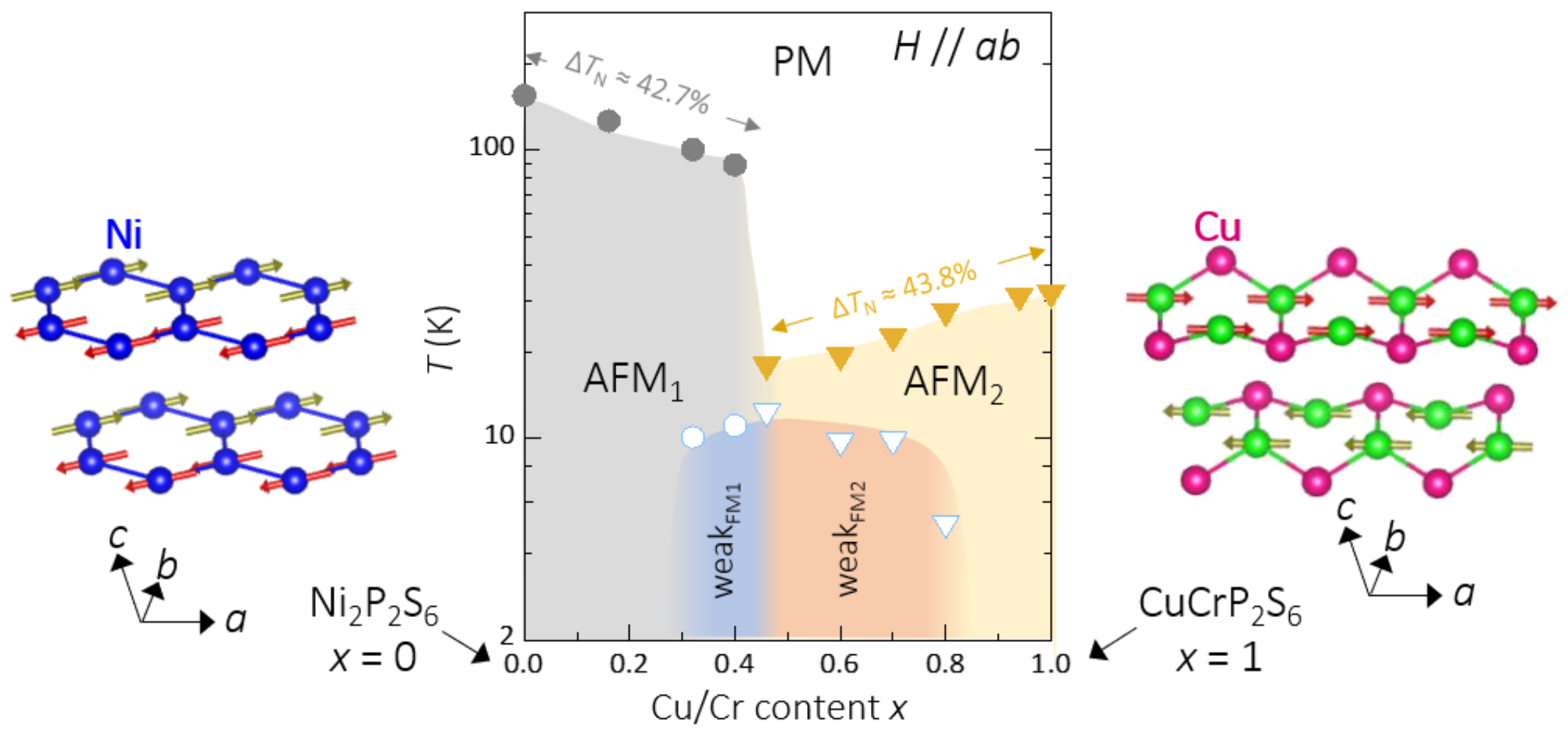

FIG. 4. The magnetic phase diagram for $Cu_xNi_{2(1-x)}Cr_xP_2S_6$ ($0 \leq x \leq 1$) showing evolution of magnetic phases with doping and temperature. The magnetic structures of pristine $Ni_2P_2S_6$ ($x$ = 0) and $CuCrP_2S_6$ ($x$ = 1) are presented on the sides of the phase diagram and denoted by $AFM_1$ (grey color) and $AFM_2$ (yellow color) in the phase diagram, respectively. The low-temperature ZFC and FC irreversibility for intermediate compositions $0.32 \leq x \leq 0.40$ and $0.46 \leq x \leq 0.80$ within the $AFM_1$ and $AFM_2$ phases depicted in Fig. 3(a) are manifested as $weak_{FM1}$ (blue color) and $weak_{FM2}$ (orange color), respectively.

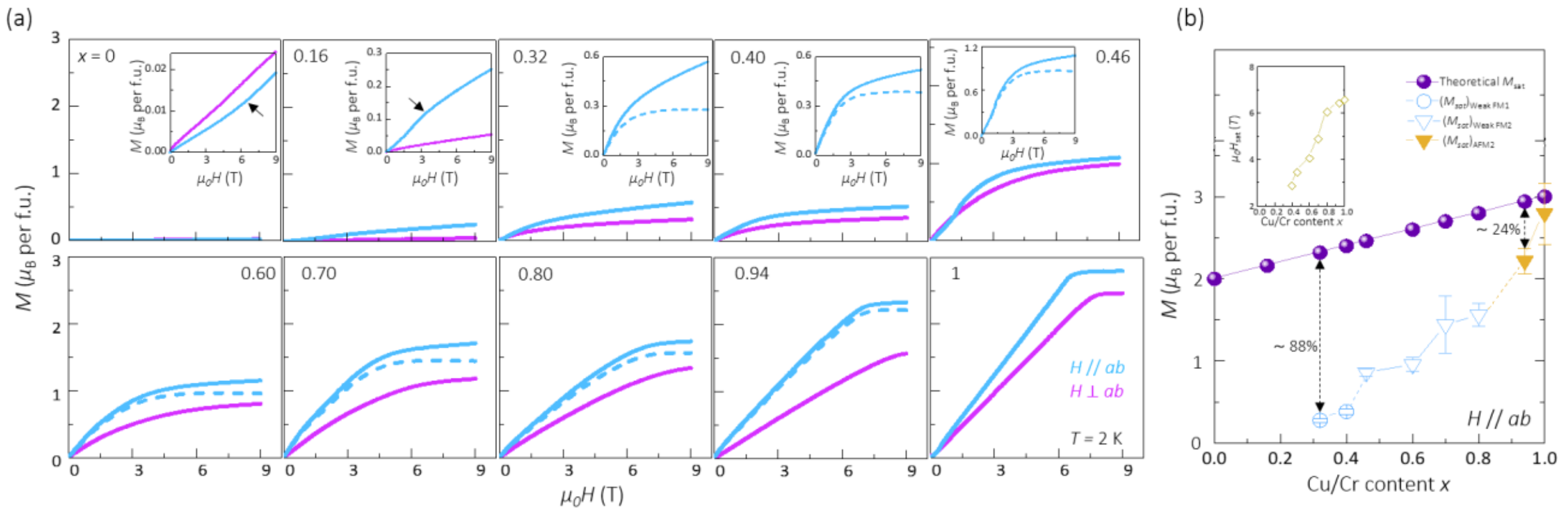

FIG. 5. (a) Field dependence of magnetization for $Cu_xNi_{2(1-x)}Cr_xP_2S_6$ ($0 \leq x \leq 1$) samples at $T = 2$ K under in-plane ($H||ab$, solid blue) and out-of-plane ($H \perp ab$, solid magenta) magnetic fields. Inset: Field-dependent magnetizations to show the metamagnetic spin-flop (SF) transitions under $H||ab$ magnetic field (denoted by black triangles for $x = 0$ and 0.16 samples) and magnetization saturation (for $x = 0.32$, 0.40, and 0.46 samples). The dashed blue lines represent the magnetization saturation after removing the linear AFM magnetization background from total in-plane magnetization (solid blue). (b) Doping dependence of saturated magnetization ($M_{sat}$) together with the expected $M_{sat}$ values. The $M_{sat}$ values for $x = 0$ and 0.16 samples are not shown because these samples lack magnetic saturation behavior up to magnetic field of $\mu_0 H = 9$ T. Inset: Doping dependence of saturation field ($\mu_0 H_{sat}$) for ferromagnetic-like polarization.

**Table I**: Nominal elemental compositions of the source materials and the actual compositions of the grown crystals determined by EDS.

| $Cu_xNi_{2(1-x)}Cr_xP_2S_6$ | |
|---|---|
| Nominal *x* | EDS *x* |
| 0 | 0 |
| 0.20 | 0.16 |
| 0.40 | 0.32 |
| 0.60 | 0.40 |
| | 0.46 |
| | 0.60 |
| 0.80 | 0.70 |
| | 0.80 |
| 0.90 | 0.94 |
| 1 | 1 |